\documentclass [12pt]{article}

\begin{document}
\title{\bf Cosmological Consequences of the Yukawa Potential Involvement in 
Gravitational and Electromagnetic Fields Interaction Evaluated by ENU Model}
\author{Miroslav S\'{u}ken\'{\i}k and Jozef \v{S}ima \\[1ex]
  Slovak Technical University, FCHPT, Radlinsk\'{e}ho 9, \\
  812 37 Bratislava, Slovakia
  \\
  e-mail:  sima@chtf.stuba.sk} \date{}
\maketitle
\begin{abstract}
  \label{sec:abstractne}
  Model of Expansive Nondecelerative Universe (ENU) due to involving
  Vaidya metric makes it possible to localize the energy density of
  gravitational field. This its capability allows for answering
  several open questions and supporting (or contradicting) results
  obtained using different theoretical approaches. This paper is aimed
  to contributing to a solution of some cosmological problems, in
  particular those linked to the Yukawa potential consequences,
  independent way of dark matter estimation, ionospheric F2 layer
  behaviour, and cyclotron radiation.
\end{abstract}

\section{Introduction}

Starting from the beginning of 80's, the inflation model of the
universe acquired dominant position in cosmology. The model has been
able to eliminate certain cosmological problems, at the same time it
has, however, open new questions, such as the Universe age, Hubble's
constant or deceleration parameter values. It has not, however,
contributed to deepen our understanding of the gravitation and its
relation to the other physical interactions. Moreover, in accordance
with some analyses [1], the initial nonhomogenities should not be
eliminated but they are rather enhanced within the inflation period.

Open questions have been a challenge for developing further models of
the Universe, one of them being Expansive Nondecelerative Universe
model [2-4].  The ENU model manifest its capability both by offering
answers to several open questions and supporting results obtained by
other theoretical approaches. The present paper continues in this
tendency and is aimed to contributing to a solution of some
cosmological problems, in particular those linked to the Yukawa
potential consequences, independent way of dark matter estimation,
ionospheric F2 layer behaviour, and cyclotron radiation.

\section{Background of Expansive Nondecelerative Universe model}

The cornerstones of ENU, rationalized in more detail in [2-4] are as 
follows: 

\begin{enumerate}
\item The Universe, throughout the whole expansive evolutionary phase,
  expands by a constant rate equals the speed of light $c$ obeying
  thus relation
\begin{equation}
\label{eq1}
r_U = c\,t_U = \frac{2\,G\,m_U }{c^2}
\end{equation}
where $r_U $ is the Universe radius, $t_U $ is the cosmological time,
$m_U $ is the Universe mass (their present ENU-based values are $r_U
\cong 1.299\,x\,10^{26}$ m, $m_U \cong 8.673\,x\, 10^{52}$ kg, $t_U
\cong 1.373\,x\,10^{10}$ yr, and match well with generally accepted
values [5-7]).

\item The curvature index and Einstein cosmologic constant are of zero
  value.
  
\item The mean energy density of the Universe is identical just to its
  critical density (its present value given by the ENU is $9.536\times
  10^{ - 27}$ kg/m$^{3}$, that provided in [6] is $9.47\times 10^{ -
    27}$ kg/m$^{3})$.

\item Since $a$ is increasing in time, $m_U $ must increase as well,
  $i.e$. in the ENU, the creation of matter occurs. The total
  mass-energy of the Universe must, however, be exactly zero. It is
  achieved by a simultaneous gravitational field creation, the energy
  of which is negative. The fundamental mass-energy conservation law
  is thus observed.

\item Due to the matter creation, Schwarzschild metrics must be
  replaced in ENU by Vaidya metrics [8, 9] in which the line element
  is formulated as
\begin{eqnarray}
ds^2 &=& \left( {\frac{d\Psi }{c\,dt}} \right)^2\frac{1}{f_{(m)}^2 }\left( {1 
- \frac{2\,\Psi }{r}} \right)\,c^2\,dt^2 - \left( {1 - \frac{2\,\Psi }{r}} 
\right)^{ - 1}dr^2 \nonumber \\
 &&- r^2\left( {d\theta ^2 + \sin ^2\theta \,d\varphi ^2} \label{eq2}
\right)
\end{eqnarray}
and the scalar curvature $R$ (which is, contrary to a more frequently used 
Schwarzschild metric, of non-zero value in Vaidya approach also outside a 
body allowing thus to localize the gravitational energy density) in the form
\begin{equation}
\label{eq3}
R = \frac{6\,G}{c^3\,r^2}.\frac{dm}{dt} = \frac{3\,r_g }{r_U \,r^2}
\end{equation}
where $m$ is the mass of a body, $G$ ($6.67259\times 10^{ - 11}$ kg$^{ - 1}$ 
m$^{3}$ s$^{ - 2})$ is the gravitational constant, $r$ is the distance from 
the body, $r_g $ is the gravitational radius of the body, $f_{(m)} $ is an 
arbitrary function, and $\Psi $is defined [3] as 
\begin{equation}
\label{eq4}
\Psi = \frac{G\,m}{c^2}
\end{equation}
In order to $f_{(m)} $ be of nonzero value, it must hold 
\begin{equation}
\label{eq5}
f_{(m)} = \Psi \left[ {\frac{d}{dr}\left( {1 - \frac{2\,\Psi }{r}} \right)} 
\right] = \frac{2\,\Psi ^2}{r^2}
\end{equation}
Based on (\ref{eq1}), in the framework of the ENU model
\begin{equation}
\label{eq6}
\frac{d\Psi }{c\,dt} = \frac{\Psi }{r_U }
\end{equation}
\end{enumerate}

Dynamic character of the ENU is described by Friedmann equations. 
Introducing dimensionless conform time, equation (\ref{eq1}) can be expressed as
\begin{equation}
\label{eq7}
c\,dt = r_U \,d\eta 
\end{equation}
from which 
\begin{equation}
\label{eq8}
r_U = \frac{dr_U }{d\eta }
\end{equation}
Applying Vaidya metric and stemming from Robertson-Walker approach, 
Friedmann equations can be then written in the form [10] 
\begin{equation}
\label{eq9}
\frac{d}{d\eta }\left( {\frac{1}{r_U }.\frac{dr_U }{d\eta }} \right) = - 
\frac{4\pi \,G}{3c^4}r_U^2 (\varepsilon + 3p)
\end{equation}
\begin{equation}
\label{eq10}
\left( {\frac{1}{r_U }.\frac{dr_U }{d\eta }} \right)^2 = \frac{8\pi 
\,G}{3c^4}r_U^2 \varepsilon - k
\end{equation}
where $\varepsilon $ is the critical energy density (the Universe actual 
density within the ENU model) and $p$ is the pressure. Based on (\ref{eq9}) and (\ref{eq10}) 
it follows
\begin{equation}
\label{eq11}
\varepsilon = \frac{3\,c^4}{8\pi \,Gr_U^2 }
\end{equation}
\begin{equation}
\label{eq12}
p = - \frac{\varepsilon }{3}
\end{equation}

Equations (\ref{eq11}) and (\ref{eq12}) represent the matter creation and the negative value 
of gravitational energy, respectively (for more details, see [2 - 4]). 

Applying Vaidya metric to one body (a system with the mass $m)$, a distance 
dependence of the gravitational field density $\varepsilon _g $ created by 
this body obeys relation (\ref{eq13}) in the weak field condition 
\begin{equation}
\label{eq13}
\varepsilon _g = - \frac{R\,c^4}{8\,\pi \,G} = - \frac{3\,m\,c^2}{4\,\pi 
\,r_U \,r^2}
\end{equation}
The absolute value of the gravitational field density of a system
equals to the critical density at the distance $r_{ef} $ (effective
interaction range of its gravitational force). Comparison of the
relations (\ref{eq4}), (\ref{eq11}) and (\ref{eq13}) leads to
\begin{equation}
\label{eq14}
r_{ef} = \left( {r_g \,r_U } \right)^{1 / 2} = \left( {2\,\Psi \,r_U } 
\right)^{1 / 2}
\end{equation}
in which $r_g $ means the gravitational radius of a body with the mass $m.$

Vaidya metric may be applied in all cases for which the gravitational energy 
is localizable, i.e. in cases being governed by relation (\ref{eq15}) 
\begin{equation}
\label{eq15}
r < r_{ef} 
\end{equation}
Gravitational influence can be thus actually realized only if the absolute 
value of the gravitational energy density created by a body exceeds the 
critical energy density, i.e. the mean gravitational energy density of the 
Universe. If
\begin{equation}
\label{eq16}
r \ge r_{ef} 
\end{equation}
Vaidya metric adopts the form of Schwarzschild metric preventing the energy 
of gravitational field from localization.

A typical feature of the ENU model lies in its simplicity, in fact that no 
``additional parameters'' or strange ``dark energy'' are needed, and in the 
usage of only one state equation in describing the Universe.

\section{Yukawa potential and our solar system}

The potential bearing Yukawa name was introduced into physics in 1935 in 
connection with prediction of mesons [11]. Originally it was formulated in 
the form 
\begin{equation}
\label{eq17}
E_{p(r)} = - \frac{E_o r_o }{r}\exp ( - r / r_o )
\end{equation}
where $E_{p(r)} $ is the nuclear potential energy between two nucleons at 
the distance $r$, $r_o $ is the range of the nuclear forces, $E_o $ is the 
strength of the interaction. A similar potential was, however, formulated 
some years earlier by Neumann and Seeliger when describing gravitational 
interactions. 

Later on, the Yukawa potential demonstrated its importance in solving 
various issues of particle physics and penetrated into chemistry, mainly in 
rationalization of liquid-vapour equilibrium, monomer-dimer mixtures 
behaviour, etc. [12--14].

Reevaluating Einstein equations, Zhuck highlighted the importance of the 
Yukawa potential for the Universe understanding and manifested its 
importance for the gravity forces in his recent paper paper [15].

In this paper the Yukawa-type gravitational potential is discussed from the 
viewpoint of its applicability for a single body and, more generally, for 
any system with defined dimensions such as a galaxy, supercluster, etc.). 
Differences emerging from application of the Newtonian and Yukawa potentials 
are exemplified by our solar system, eliptic galaxies and super clusters 
parameters. The discussion is held in the framework of the Expansive 
Nondecelerative Universe model (ENU). 

Taking Vaidya metric into account, the non-relativistic gravitational 
potential outside a body with the mass $m$ can be expressed in the form
\begin{equation}
\label{eq18}
\Phi = \Phi _o \exp ( - r / r_{ef} )
\end{equation}
where
\begin{equation}
\label{eq19}
\Phi _o = - G\int {\frac{\rho \,}{r}dV} = - \frac{G\,m}{r}
\end{equation}
It follows both from the Zhuck [15] and our approaches that consequences of 
the potential, i.e. differences between results provided using the Newtonian 
and Yukawa potentials can actually be observed only taking sufficiently 
large distances $r$ into account. We have chosen our solar system (its the 
most distant planets - Neptun and Pluto), and eliptic galaxies as examples. 
The results obtained enable us to estimate the amount of dark matter in the 
investigated galaxies and in the whole Universe.

The Sun mass is [16] 
\begin{equation}
m_{(Sun)} = 1.9891\times 10^{30} \mbox{kg}
\end{equation}
which represents 99.866{\%} of the mass of the whole solar system,
$m_{SS}$, being
\begin{equation}
m_{(SS)} \cong 1.9918\times 10^{30} \mbox{kg}
\end{equation}
The effective interaction range of the solar system $r_{ef(SS)} $ based on 
(\ref{eq14}) and (21) is as follows
\begin{equation}
r_{ef(SS)} \cong 6.43\times 10^{14} \mbox{m}
\end{equation}
A mean square of velocity of the planets revolution can be expressed, the 
Yukawa potential including, as 
\begin{equation}
\label{eq20}
\bar {v}^2 = \frac{4\,\pi ^2\,\bar {r}^2}{t^2} = \frac{Gm_{(SS)} }{\bar 
{r}}\exp ( - \bar {r} / r_{ef} )
\end{equation}
of which
\begin{equation}
\label{eq21}
 - r_{ef} \ln \frac{4\pi ^2\,\bar {r}^3}{G\,m_{(SS)} \,t^2} = \bar {r}
\end{equation}
where $t$ is the orbital period of a planet around the Sun, $\bar {r}$ is a 
mean Sun-to-planet distance. Relations (21), (22), (\ref{eq21}), and known $t$ for 
Neptun lead to the following ENU-based value
\begin{equation}
\bar {r}_{(Neptun)ENU} = 4.488\times 10^{12} \mbox{m}
\end{equation}
Calculation based on the Newtonian potential gives
\begin{equation}
\bar {r}_{(Neptun)NP} = 4.496\times 10^{12} \mbox{m}
\end{equation}
Difference of both the values is about 0.2{\%}. A similar treatment for 
Pluto leads to
\begin{equation}
\bar {r}_{(Pluto)ENU} = 5.862\times 10^{12} \mbox{m}
\end{equation}
and
\begin{equation}
\bar {r}_{(Pluto)NP} = 5.878\times 10^{12} \mbox{m}
\end{equation}
i.e. about 0.3{\%} deviation.

\section{The Universe dark matter estimation}

Generally, the more massive gravitationally bonded structure, the lower its 
energy density. It can be supposed that for superclusters or giant clusters 
of galaxies, their energy density approaches the critical density. It means 
that dimension of such systems is
\begin{equation}
\label{eq22}
r \cong r_{ef} 
\end{equation}
Consequently, the total mass of such systems dark, matter including, may be 
determined directly from their dimensions. Based on (\ref{eq14}) and (\ref{eq22}) it follows
\begin{equation}
\label{eq23}
m_{(Galaxy,total)} \cong \frac{r^2\,c^2}{2\,G\,r_U }
\end{equation}
The observable mass of the most massive, up-to-now known eliptic galaxies 
reaches
\begin{equation}
\label{eq24}
m_{(Galaxy,obs)} \approx 10^{13}m_{(Sun)} 
\end{equation}
and their dimensions are about
\begin{equation}
r \cong r_{ef} \cong 1.56\times 10^{22}\mbox{m}
\end{equation}
Stemming from (\ref{eq23}) and (32) the total mass of such galaxies, dark matter 
including, represents
\begin{equation}
\label{eq25}
m_{(Galaxy,total)} \approx 10^{15}m_{(Sun)} 
\end{equation}
An observable mass of an average-sized super cluster is
\begin{equation}
\label{eq26}
m_{(SC,obs)} \approx 10^{16}m_{(Sun)} 
\end{equation}
and its dimension is about
\begin{equation}
r_{(SC)} \cong r_{ef} \cong 7.1\times 10^{23} \mbox{m}
\end{equation}

Based on (\ref{eq23}), the total mass of a super cluster reaches
\begin{equation}
\label{eq27}
m_{(SC,total)} \approx 10^{18}m_{(Sun)} 
\end{equation}
which directly leads, comparing the values mentioned in (\ref{eq25})
and (\ref{eq27}), to a result that the total mass exceeds by about two
orders the "visible" mass, i.e. the majority of the total matter is
hidden in the form of dark matter.

Based on validity of (\ref{eq22}), relations (\ref{eq20}) and
(\ref{eq23}) provide the rotational velocity of super cluster (SC) and
giant eliptic galaxies (EG)
\begin{equation}
\label{eq28}
v = c\left( {\frac{r}{2\times 2.71\,\,r_U }} \right)^{1 / 2}
\end{equation}
Provided that the dimensions (radius) of gravitationally bonded systems - 
super clusters and giant eliptic galaxies - are known, and including the 
gauge factor $r_U $ value into account, values of their rotational velocity 
emerge 
\begin{eqnarray}
v_{(EG)} &\cong& 1.3\times 10^6 \mbox{m/s}\\
v_{(SC)} &\cong& 8.9\times 10^6  \mbox{m/s}
\end{eqnarray}
The above mentioned ENU-based calculated values are in good accordance with 
those experimentally observed.

\section{Breaking electron-cation radiation of the F2 layer of ionosphere}

The F2 layer is part of the Earth ionosphere formed predominantly by
atomic ogyxen cations and electrons (the concentration and kind of
neutral molecules, a degree of ionization and temperature depend on
several factors, the most important being solar radiation intensity,
season of the year, etc.) [17, 18]. Our attention will be focused on
the electrons present in the F2 layer.

The effective cross section $d\sigma $ related to the emission of a
photon (brems\-strahlung, i.e. breaking radiation) in the frequency
interval $d(\hbar \omega )$ by an electron in the field of a cation
with the charge $Z\,e$ is defined by Bethe-Heitler formula [19, 20]
\begin{equation}
d\sigma = \frac{8Z^2\alpha \,\,r_e^2 }{3}.\frac{m_e c^2}{E_e }.\frac{d\omega 
}{\omega }.\ln \frac{\left[ {E_e^{1 / 2} + \left( {E_e - \hbar \omega } 
\right)^{1 / 2}} \right]^2}{\hbar \omega }
\end{equation}
where $\alpha $ is the fine structure constant, $m_e ,\;r_e ,\,E_e $ are the 
electron mass, radius, and kinetic energy, respectively. 

The radiation output of a plasma volume unit is obtained by multiplying the 
photon energy $\hbar \omega $ with the current density of electrons $n_e $ 
colliding with cations having the density $n_i .$ In a limiting case of the 
total one-electron ionization (a state not far from the actual state 
approached at favourite atmospheric conditions), $Z = 1,\;n_e = n_i = n$, 
and
\begin{equation}
dI_{(\omega )} = n^2\hbar \omega \,v\,d\sigma 
\end{equation}
At integration of (40) the following substitution is applied
\begin{equation}
2\,E_e = \hbar \omega \left( {1 + chx} \right)
\end{equation}
Then, based on (40) to (42), the total radiation output of a plasma
volume unit at the electron-cation collisions is given as
\begin{equation}
I_{(e_i )} = \frac{32}{3}.\left( {\frac{2}{\pi }} \right)^{1 / 2}\alpha 
\,c^3r_e^2 m_e n^2t^{1 / 2}
\end{equation}
where
\begin{equation}
t = \frac{k\,T}{m_e c^2}
\end{equation}
It is generally accepted that the F2 layer peaks at about $2\times 10^{12}$ 
e$^{ - }$/m$^{3}$ during the day and $5\times 10^{10}$e$^{ - }$/m$^{3}$ 
during the night. The electron temperature is widely spread in the range of 
1000 K to 3000 K. On the other hands, for the cations, the temperature 
covers 800 K to 1500 K. The pressure in this region ranges between $10^{ - 
5} - 10^{ - 6}$ Pa and the ionic mean free path is 4.5 to 15 km. 
Accordingly, the ionospheric plasma can be considered as a highly ionized, 
almost collisionless gas in the F2 region [18]. Based on the above values 
and relation (43), the total output of the breaking radiation in a volume 
unit of the F2 layer reaches (in the calculation, limiting conditions of the 
temperature interval from 1000 K to 3000 K, and electron concentration from 
$1\times 10^{12}$ e$^{ - }$/m$^{3}$ to $2\times 10^{12}$ e$^{ - }$/m$^{3}$ 
were taken) 
\begin{equation}
I_{(e_i )} = 4.9\times 10^{ - 15} - 3.3\times 10^{ - 14} \mbox{W/m}
\end{equation}
Gravitational output is in the ENU defined as 
\begin{equation}
\left| {P_g } \right| = \frac{d}{dt}\int {\frac{c^4}{8\,\pi \,G}} \,R\,dV = 
\frac{m\,c^3}{r_U }
\end{equation}
Taking the concentration of oxygen cations in a volume unit of the F2 layer, 
\begin{equation}
\left| {P_g } \right| = 5.0\times 10^{ - 15} - 1.0\times 10^{ - 
14} \mbox{W/m}
\end{equation}
Thus, during the daytime, the gravitational output approaches the breaking 
radiation output (intervals of the outputs overlaps, as shown by (45) and 
(47)).

\section{Cyclotron radiation of the Sun surface}

The mass density at the Sun surface is about
\begin{equation}
\rho \cong 1 \mbox{kg/m$^{3}$}
\end{equation}
which corresponds to particle concentration
\begin{equation}
n \approx 10^{27} \mbox{m$^{-3}$}
\end{equation}
at the Sun surface temperature
\begin{equation}
T \cong 5.8\times 10^3 \mbox{K}
\end{equation}
At these conditions (further, for calculation purposes, the value expressed 
in (48) is taken as an exact value), the ionization degree is
\begin{equation}
\chi = 1.161\times 10^{ - 6}
\end{equation}
which leads to the concentration of free electrons
\begin{equation}
n_{(e)} = 1.161\times 10^{21} \mbox{m$^{ - 3}$}
\end{equation}
At such conditions the Debye-H\"{u}ckel radius is still higher than
the mean ion-ion distance, i.e. the kinetic energy of the present
particles exceeds electromagnetic Coulomb energy of the ions.

Owing to the relatively great mass of the present cations, their
cyclotron radiation may be omitted. Breaking electron-cation and
electron-electron radiation (bremsstrahlung [19]) is due to a high
cation concentration extremely high.

A particle with the charge $e$ and acceleration $a$ emits in a second
the energy
\begin{equation}
\label{eq2a}
E_c = a^2.\frac{e^2}{6\,\pi \,\varepsilon _{o\,} c^3}
\end{equation}
Following substitutions
\begin{equation}
\label{eq3a}
a = \frac{e\,\,v_e \,B}{m_e }
\end{equation}
and
\begin{equation}
\label{eq4a}
r_e = \frac{e^2}{4\,\pi \;\varepsilon _o m_e c^2}
\end{equation}
equation (\ref{eq2a}) may be expressed as 
\begin{equation}
\label{eq5a}
E_c = \frac{8\,\pi \,r_e^2 c\,\varepsilon _o v_e^2 B^2}{3}
\end{equation}
where
\begin{equation}
\label{eq6a}
v_e^2 = \frac{2\,kT}{m_e }
\end{equation}

The total cyclotron radiation in a volume unit of the plasma forming the Sun 
surface is
\begin{equation}
\label{eq7a}
I_c = n_e E_c = \frac{16\,\pi \,r_e^2 c\,\varepsilon _o kT\,n_e B^2}{3\,m_e 
}
\end{equation}
The gravitational output of this volume unit is
\begin{equation}
P_g = \frac{\rho _{pl} \,c^3}{r_U } \cong 0.2 \mbox{W/m$^{3}$}
\end{equation}
provided that the plasma density is
\begin{equation}
\rho _{pl} \cong 1 \mbox{kg/m$^{3}$} 
\end{equation}
The identity of (\ref{eq7a}) and (59) gives the value of induction
\begin{equation}
B \cong 7.6\times 10^{ - 2} \mbox{T}
\end{equation}
This is a real value since the maximal value at Sun spots approaches 
\begin{equation}
B_{\max } \cong 10^{ - 1} \mbox{T}
\end{equation}
The magnetic field energy density is defined as
\begin{equation}
\label{eq8a}
W_B = \frac{B^2}{2\,\mu _o }
\end{equation}
Then, a comparison of (13) and (\ref{eq8a}) leads to the critical Sun induction 
value
\begin{equation}
B_{crit} \cong 3.8\times 10^{ - 2} \mbox{T}
\end{equation}
The above results, namely very closed values of (61) and (64) allow us to 
formulate a conclusion stating that the cyclotron and gravitational output 
of a volume unit of the plasma forming the Sun surface are (almost) 
identical which may have an effect on the Sun corona heating.

\section{Cyclotron radiation at the end of radiation era}

At the end of radiation era, the concentration of ions was very low
and, consequently, the effect of breaking electron-ion and
electron-electron radiation may be omitted. This part will be devoted
and limited to cyclotron radiation only. Stemming from the identity of
(\ref{eq5a}) and (\ref{eq13a}), introduction of the expressions
(\ref{eq9a}) -- (\ref{eq12a})
\begin{equation}
\label{eq9a}
r_e = \frac{\alpha \,\hbar }{m_e c}
\end{equation}
\begin{equation}
\label{eq10a}
v_e^2 = \frac{2\,kT}{m_e }
\end{equation}
\begin{equation}
\label{eq11a}
kT = E_{Pc} .\left( {\frac{l_{Pc} }{r_U }} \right)^{1 / 2}
\end{equation}
\begin{equation}
\label{eq12a}
B^2 = \frac{3\,c^2}{8\,\pi \,G\,r_U^2 \varepsilon _o }
\end{equation}
and application of the gravitational output of an electron
\begin{equation}
\label{eq13a}
\left| {P_e } \right| = \frac{d}{dt}\int {\frac{c^4}{8\,\pi \,G}} .R\,dV = 
\frac{m_e c^3}{r_U }
\end{equation}
lead to
\begin{equation}
r_{U(re)} = \left[ {\frac{2\;\alpha ^2E_{Pc} \,\hbar ^2\left( {l_{Pc} } 
\right)^{1 / 2}}{m_e^4 G\,c^2}} \right]^{2 / 3} \approx 10^{22} \mbox{m} 
\end{equation}
where $E_{Pc} (1.2211\times 10^{19}$ eV) and $l_{Pc} (1.616051\times 10^{ - 
35}$ m) are the Planck energy and length, respectively. Relation (70) shows 
that at the end of radiation era, the electron emitted identical amounts of 
the cyclotron radiation and gravitational energy. It should be pointed out 
that the gravitational influence of the electron was not observable at that 
time.

The Compton wavelength of a proton is
\begin{equation}
\label{eq14a}
\lambda _p = \frac{\hbar }{m_p c}
\end{equation}
where $m_p $ is the proton mass. When comparing (14) and (\ref{eq14a}), the time of 
the proton gravitational influence on its surrounding is obtained. It is the 
time in which the Universe radius reached the value 

\begin{equation}
r_{U(re)} = \frac{\hbar ^2}{2\,G\,m_p^3 } \approx 10^{22} \mbox{m}
\end{equation}
i.e. just at the end of radiation era. Putting relations (70) and (72) 
identical, a formula relating the proton and electron masses is obtained in 
the form 
\begin{equation}
\label{eq15a}
\frac{m_p }{m_e } = \left( {\frac{m_{Pc} }{32\,m_p }} \right)^{1 / 8}\alpha 
^{ - 1 / 2} = 1860
\end{equation}
The value in (\ref{eq15a}) is in excellent agreement with the experimental value.

\subsection{Acknowledgement}

This work was supported by the Grant Agency VEGA through grant No. 1/9251/2.

\section*{References}
\begin{description}
\item [] [1] R. Penrose, The Large, the Small and the Human Mind, Cambridge University 
Press, Cambridge, 1997, p. 44

\item [] [2] V. Skalsk\'y, M. S\'uken\'{\i}k, Astroph. Space Sci., 178 (1991) 169; 181 (1991), 
153; 191 (1992) 329; 209 (1993) 123; 215 (1994) 137;

\item [] [3] M. S\'uken\'{\i}k, J. \v{S}ima, Spacetime and Substance, 3 (2001) 125

\item [] [4] J. \v{S}ima, M. S\'uken\'{\i}k, Entropy, 4 (2002) 152

\item [] [5] R. Cayrel, V. Hill, T.C. Boers, B. Barbuy, M. Spite, F. Spite, B. Plez, 
J. Anderson, P. Bonifacio, P. Francois, P. Molaro, B. Nordsrom, F. Primas, 
Nature 409 (2001) 691 

\item [] [6] D.E. Groom et al., Eur. Phys. J., C15 (2000) 1 

\item [] [7] A.H. Guth, The Inflationary Universe, Addison Wesley, New York, 1997, 
p.22

\item [] [8] P.C. Vaidya, Proc. Indian Acad. Sci., A33 (1951) 264

\item [] [9] K.S. Virbhadra, Pramana -- J. Phys., 38 (1992) 31

\item [] [10] A.N. Monin, Cosmology, Hydrodynamics and Turbulence: A.A. Friedmann and 
Extension of jis Scientific Heritage, Moscow, 1989, p.103 (in Russian)

\item [] [11] H. Yukawa, Progr. Theor. Phys. 17 (1935) 48
\item [] [12] N. Sulaiman, S.M. Osman, I. Ali, R.N. Singh, J. Non-Crystalline Solids 
312 (2002) 227
\item [] [13] S.M. Osman, I. Ali, R.N. Singh, J. Phys.: Condens. Matter 14 (2002) 8415
\item [] [14] I. Ali, S.M. Osman, M. Al-Busaidi, R.N. Singh, Int. J. Mod. Phys., B13 
(1999) 3261
\item [] [15] N.A. Zhuck, Spacetime and Substance, 2 (2001) 154
\item [] [16] Handbook of Chemistry and Physics, 78$^{th}$ Ed., CCR, Boca Raton (1997) 
\item [] [17] M.C. Kelley, The Earth's Ionosphere, Academic Press, San Francisco, 1989
\item [] [18] H-J. Rhee, J. Astron. Space Sci., 14 (1997) 269
\item [] [19] H.A. Bethe, W. Heitler, Proc. Royal Soc., London, A146 (1934) 83
\item [] [20] W. Heitler, Quantum Theory of Radiation, Clarendon Press, Oxford, 1954
\end{description}

\end{document}